# Criticality in Cell Differentiation

Indrani Bose and Mainak Pal

Department of Physics

Bose Institute

93/1, A. P. C. Road

Kolkata-700009, India

Abstract. Cell differentiation is an important process in living organisms. Differentiation is mostly based on binary decisions with the progenitor cells choosing between two specific lineages. The differentiation dynamics have both deterministic and stochastic components. Several theoretical studies suggest that cell differentiation is a bifurcation phenomenon, well-known in dynamical systems theory. The bifurcation point has the character of a critical point with the system dynamics exhibiting specific features in its vicinity. These include the critical slowing down, rising variance and lag-1 autocorrelation function, strong correlations between the fluctuations of key variables and non-Gaussianity in the distribution of fluctuations. Recent experimental studies provide considerable support to the idea of criticality in cell differentiation and in other biological processes like the development of the fruit fly embryo. In this Review, an elementary introduction is given to the concept of criticality in cell differentiation. The correspondence between the signatures of criticality and experimental observations on blood cell differentiation in mice is further highlighted.

Cell differentiation is the process through which stem or progenitor cells diversify into different cell types such as kidney, liver and skin cells. Cells, in general, have identical sets of genes. The different cell types are distinguished by distinct gene expression profiles. A gene active in one cell type may be either silent or expressed at a lower level in another cell type. Three representations which capture well the broad aspects of cell differentiation are Waddington's epigenetic landscape [1], the mammalian cell-fate tree [2] and the potential landscape [3]. In Waddington's landscape, a single marble, representing a cell, rolls down a cascade of branching valleys separated by hills. Each branching is binary in nature, i.e., the marble can slide into one of two valleys. The terminating valleys of the cascade represent stable cell types. In the mammalian cell-fate tree, the tree starts from a single root depicting the embryonic stem cell and has a binary branching structure. At each branch point, the cell has a choice between two different lineages with the terminal branches of the tree associated with stable, distinct cell types. In the potential landscape, the potential function (to be defined) is plotted in the space of all gene expression states, termed the state space. The landscape consists of hills and valleys with the valleys and hilltops describing stable differentiated and metastable progenitor cell states respectively. A vast body of experimental knowledge on cell differentiation provides the basis for the construction of quantitative theoretical models to gain insight on the physical principles underlying the differentiation process. The mathematical and computational formalisms used to elucidate the principles utilize the concepts and techniques of nonlinear dynamics and the theory of stochastic processes. In the Glossary, we provide a definition of some basic terminologies of nonlinear dynamics.

Kauffman [4,5] carried out an exhaustive simulation of the dynamics of a genetic network and put forward the hypothesis that the different cell types correspond to the attractors of the dynamics of a network of thousands of interacting genes. The experimental evidence that a high-dimensional stable attractor describes a distinct cellular phenotype was first provided by Huang et al [6]. A series of theoretical studies have viewed cell differentiation as a bifurcation phenomenon [1,7,8,9]. While bifurcation can be of different types, the supercritical pitchfork bifurcation provides a simple metaphor for understanding cell differentiation as a binary decision-making process. In this type of bifurcation, the stable undifferentiated cell state (monostability) differentiates into two distinct stable cell types (bistability) at a critical parameter value known as the bifurcation point. This point is characterised by the loss of stability of the monostable state and the appearance of a new pair of stable steady states. A sequence of successive pitchfork bifurcations is representative of the mammalian cell fate tree consisting of binary branches [10]. A two-gene motif (figure 1(i)) provides a simple physical picture of the cell differentiation process. The evidence for the existence of the motif, as a component of the gene regulatory networks governing cell differentiation in different cell lines, has been obtained through an exhaustive analysis of the human expression data involving 2602 transcriptionally-regulated genes and 166 distinct cell types [11]. The motif consists of the genes $X_1$ and $X_2$ the protein products of which repress each other's expression. The feedback loop, based on mutual antagonism, is equivalent to a positive feedback loop. The proteins also autoactivate their own production through individual positive feedback loops. Positive feedback loops coupled with sufficiently nonlinear dynamics promote bistability in specific parameter regimes. The simple genetic circuit captures the binary decision process at a branch point of the cell-fate tree (figure 1(ii)). In the undifferentiated progenitor state, the protein concentrations $x_1$ and $x_2$ are nearly equal whereas the stable steady states, $x_1 \gg x_2$ and $x_1 \ll x_2$, represent the differentiated cell states.

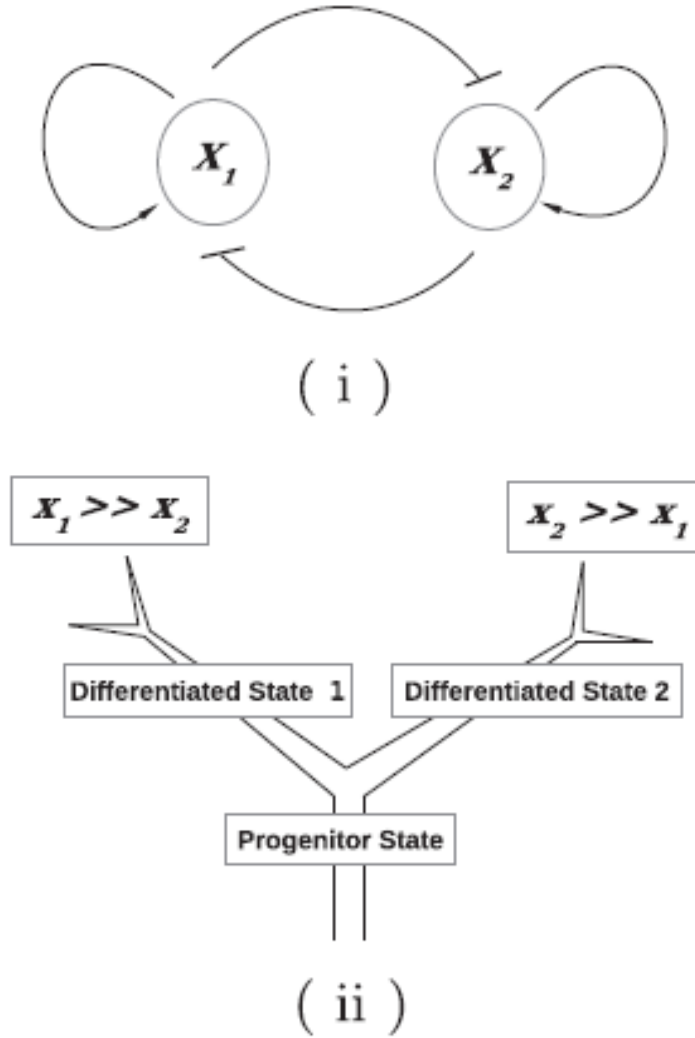

Figure 1. (i) The central motif of gene regulatory networks controlling cell differentiation. The protein products of two genes $X_1$ and $X_2$ repress each other's synthesis, thus constituting a positive feedback loop. The proteins further autoactivate their own synthesis. (ii) Illustration of a progenitor state differentiating into two states 1 and 2 distinguished by the distribution of the proteins $x_1 \gg x_2$ and $x_2 \gg x_1$.

The rate equations describing the dynamics of the two-gene motif can be written as [12]

$$\frac{dx_1}{dt} = \frac{a_1 x_1}{S + x_1} + \frac{b_1 S}{S + x_2} - k_1 x_1 - g\, x_1 x_2 \qquad (1)$$

$$\frac{dx_2}{dt} = \frac{a_2 x_2}{S + x_2} + \frac{b_2 S}{S + x_1} - k_2 x_2 - g\, x_1 x_2 \qquad (2)$$

The first terms on the right hand sides of the two equations correspond to autoactivation, the second terms represent the cross-repression of the expression of the two genes, the third terms denote the individual protein degradation rates and the fourth terms represent the indirect repression of gene expression via heterodimer formation. Considering the symmetric situation $a_1 = a_2 = a$, $b_1 = b_2 = b$ and $k_1 = k_2 = k$, one can compute the bifurcation diagram for the steady state values of $x_1$ ($x_2$) for any one of the bifurcation parameters $a$, $b$, $k$ and $g$. Figure 2 shows the bifurcation diagram, a supercritical pitchfork bifurcation, with $a$ serving as the bifurcation parameter. The other parameter values are $b = 2.0$, $k = 1.0$, $S = 2.0$ and $g = 0.5$. The solid lines represent stable steady states whereas the dotted line describes a branch of unstable steady states.

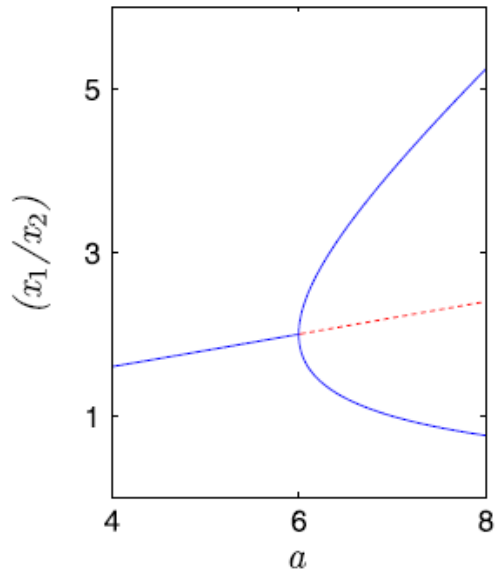

Figure 2. Bifurcation diagram depicting the steady state values $x_1$ ($x_2$) versus the bifurcation parameter $a$.

If cell differentiation were totally controlled by deterministic dynamics, the stable steady states of the cells in an undifferentiated cell population, prepared in the same initial state, would be identical and described by a single gene expression level. Recent experimental observations, however, offer evidence for considerable non-genetic heterogeneity in the gene expression profiles of both embryonic and adult stem cells [13,14]. The heterogeneity implies that instead of a single gene expression, say, protein level, the undifferentiated cell population is characterised by a distribution of protein levels. The heterogeneity arises due to the randomness associated with key biological processes like gene expression [15,16]. In the presence of stochasticity, the certainty and the preciseness of a deterministic dynamical description are lost. Stochasticity introduces noise (random fluctuations) in gene expression levels which can be quantified using single cell and single molecule methodologies [17]. In the case of a limited amount of noise, the broad features of deterministic dynamics are still valid. The state of a cell population at time $t$ is no longer represented by a single point in state space but by a cloud of points. The attractor of the dynamics, instead of being a single fixed point, has a fuzzy character in the state space. Stochastic dynamics require a statistical mechanical perspective for the appropriate description of the process of cell differentiation [18,19]. The microstate at time $t$ of an ideal gas in a closed container is described in terms of the instantaneous positions and momenta of the individual gas molecules. The macrostate of the system is characterised by average properties like pressure and temperature with numerous interchangeable microstates realizing the same macrostate. The pluripotent state of a progenitor cell population, which is yet to differentiate into daughter lineages, is not a single state but a macrostate arising out of a large number of microstates. The individual cells in the population are the analogs of gas molecules and the protein concentrations the analogs of positions and momenta. The microstate at time $t$ is defined by the sets of instantaneous protein concentrations in individual cells. The macrostate is specified by the steady state probability distribution of the protein levels in the cell population. In the case of the two-gene motif, the steady state probability distribution is given by $P_{st}(x_1, x_2)$ and a potential function may be computed as $U(x_1,x_2) = -\ln P_{st}(x_1, x_2)$. The potential landscape, mentioned in the Introduction, is obtained by plotting $U(x_1,x_2)$ in the two-dimensional (2d) state space with its two axes denoting the possible values of $x_1$ and $x_2$. In the case of a one-variable system described by the deterministic rate equation, $\frac{dx}{dt} = f(x)$, the potential function $U(x)$ is given by $f(x) = -\frac{dU}{dx}$, i.e., $U(x)$ can be computed by integrating $f(x)$ over $x$ in the appropriate range of $x$ values. This method of computing the potential function cannot, however, be generalized to the case of a multi-variable system [20]. In the stochastic description, the valleys of the potential landscape are the regions in state space in which the steady state probability distribution is maximal. In the case of deterministic dynamics, each stable steady state has its own basin of attraction and the trajectories are confined to evolve within their specific basins. With the introduction of stochasticity in the dynamics, fluctuation-driven transitions of the trajectories between different basins are possible so that a cell may acquire a fate different from the one expected on the basis of deterministic dynamics.

One simple method of investigating stochastic time evolution is to make use of the formalism based on the Langevin equations [21,22]. The Langevin equation is a rate equation with a noise term added to the usual deterministic part so that it reduces to the deterministic rate equation in the absence of noise. The steady state probability distributions of the two-gene motif, with the deterministic components of the dynamics given by equations (1) and (2), have been computed for different values of the bifurcation parameter $a$ using the formalism of the Langevin equations [12]. Figure 3(i) depicts the steady state probability distribution, $P_{st}(x_1, x_2)$, which corresponds to the macrostate of a pluripotent cell population. The broad heterogeneity in the distribution of protein levels is consistent with the experimental observations of Chang et al. [23]. In the experiment, a clonal population of mouse hematopoietic (blood-cell forming) progenitor cells was found to exhibit a nearly 1000-fold range in the levels of a stem-cell-surface marker protein Sca-1. A two-gene motif plays a key role in the differentiation process with the two antagonistic genes synthesizing the proteins GATA1 (concentration $x_1$) and PU.1 (concentration $x_2$). The cells with low (high) Sca-1 levels were found to have high (low) GATA1 and low (high) PU.1 levels. Figs. 3(ii-iv) display the steady state probability distributions after bifurcation has occurred. The initial states of the cell population have been chosen from three distinct parts of the distribution in figure 3(i). The initial states chosen are $x_1 = x_2$ (figure 3(ii)), $x_2 > x_1$ (figure 3(iii)) and $x_1 > x_2$ (figure 3 (iv)). The smaller subpopulations of differentiated cells in figures 3(iii) and 3(iv) are associated with cell fates distinct from those in the case of deterministic dynamics. The origin of these subpopulations lies in noise-induced transitions across the boundaries of the basins of attraction. The computational results are in conformity with the experimentally observed feature of multilineage priming [23] of the pluripotent cell population, i.e., the different parts of the cell population having distinct propensities for lineage choice. A recent experimental study [24] has detected the presence of "rebellious cells" which acquire the fate opposite to the one expected. A possible origin of the existence of such cells lies in noise-induced inter-basin transitions. Bifurcation can be described as a dynamical phase transition separating two distinct dynamical regimes with the bifurcation point serving as a critical point. More familiar examples of phase transitions include thermodynamic transitions like liquid-gas, paramagnet-ferromagnet and metal-superconductor transitions. For temperatures below a critical temperature, a magnetic system is in the ordered ferromagnetic phase whereas above the critical temperature the system is in the magnetically disordered paramagnetic phase. Thermodynamic phase transitions involve equilibrium phases of the system whereas bifurcation describes transitions between regimes defined in terms of nonequilibrium steady states. From a statistical mechanical perspective, the cell-fate decisions of embryonic stem cells have been likened to phase transitions involving the appearance of an appropriately defined order parameter below the transition point [18]. A feature common to both thermodynamic critical point transitions and bifurcations is that of critical slowing down.

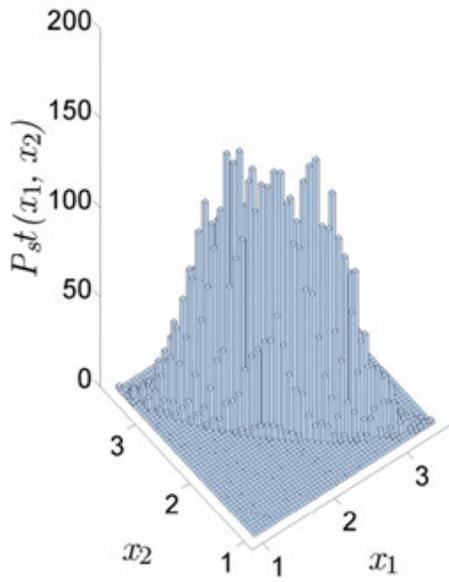
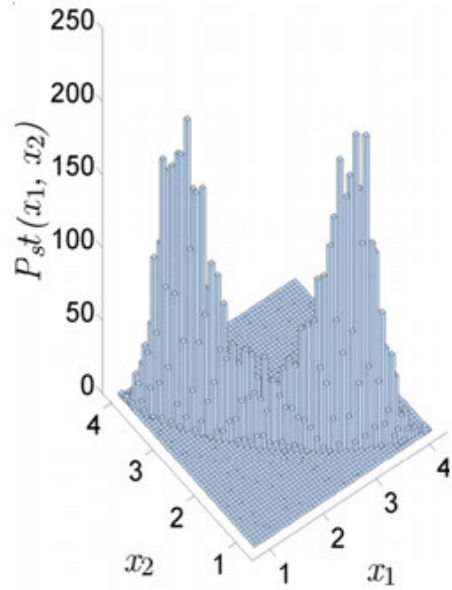

(i)

(ii)

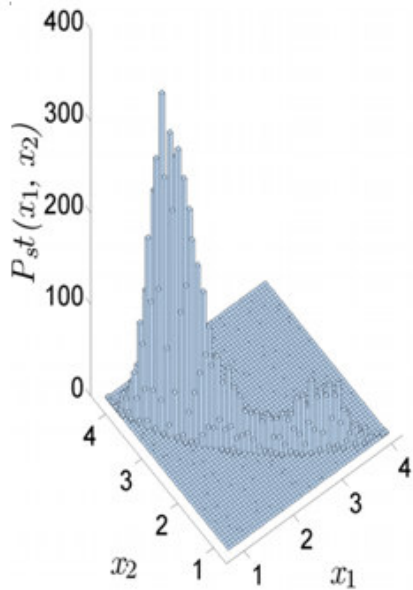
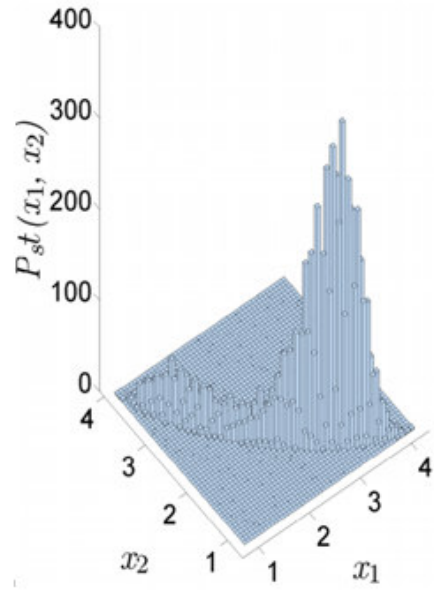

(iii)

(iv)

Figure 3. Steady state probability distributions $P_{st}(x_1, x_2)$ obtained by simulating the Langevin equations [12] containing only additive noise. (i) The parental histogram of the protein levels $x_1$ and $x_2$ in the case of undifferentiated cells. (ii)-(iv) Steady state probability distribution for differentiated cells. The forms of the distributions are dictated by the locations of the initial states in the parental distribution. The parameter $a$ serves as the bifurcation parameter.

Consider a thermodynamic (dynamic) system to be weakly perturbed so that the system moves away from the thermodynamic equilibrium state (stable steady state). The return time $T_R$, defined to be the average time taken by the system to return to the equilibrium (stable steady) state, diverges as the critical point is approached, i.e., the relaxation kinetics become extremely sluggish. We illustrate the origin of critical slowing down close to a bifurcation point in the case of the two-gene motif (Fig. 1(i)), a key component in the gene regulatory network controlling differentiation in various cell lines. The dynamics of the two-gene motif are described by the differential rate equations

$$\frac{dx_1}{dt} = f(x_1, x_2), \quad \frac{dx_2}{dt} = g(x_1, x_2) \quad (3)$$

where the functions $f(x_1, x_2)$ and $g(x_1, x_2)$ are given by the expressions on the right hand sides of equations (1) and (2) respectively. Let a steady state solution of the rate equations be denoted by the fixed points $(x_1^S, x_2^S)$. If the system is perturbed away from the steady state, the deviations of the variables from their fixed point values are represented as

$$\delta x_1 = x_1 - x_1^S, \quad \delta x_2 = x_2 - x_2^S \quad (4)$$

Define the column vector

$$\delta x = \begin{pmatrix} \delta x_1 \\ \delta x_2 \end{pmatrix} \quad (5)$$

The functions $f(x_1, x_2)$ and $g(x_1, x_2)$ are now Taylor expanded around the steady state values of the variables $x_1$ and $x_2$. In the case of weak perturbation, one needs to retain terms up to only the first order and the rate equations become

$$\frac{d(\delta x)}{dt} = J \, \delta x \quad (6)$$

where $J$ is the Jacobian matrix given by

$$\begin{pmatrix} \frac{\partial f(x_1, x_2)}{\partial x_1} & \frac{\partial f(x_1, x_2)}{\partial x_2} \\ \frac{\partial g(x_1, x_2)}{\partial x_1} & \frac{\partial g(x_1, x_2)}{\partial x_2} \end{pmatrix}_{SS} = \begin{bmatrix} -\beta_1 & r_{12} \\ r_{21} & -\beta_2 \end{bmatrix} \quad (7)$$

The diagonal elements of the Jacobian matrix represent the effective decay rates of the two proteins and the off-diagonal terms describe the repressive effects of the proteins on each other's synthesis. Let the eigenvalues of $J$ be $\lambda_1$ and $\lambda_2$ with the corresponding eigenvectors $V_1$ and $V_2$. The general solution of equation (6) is of the form

$$\delta x(t) = C_1 V_1 e^{\lambda_1 t} + C_2 V_2 e^{\lambda_2 t} \qquad (8)$$

where the coefficients $C_1$ and $C_2$ are determined by the initial conditions. The stability of the steady states requires that the real parts of $\lambda_1$ and $\lambda_2$ are negative. Let $\lambda_1$ be the largest (dominant) eigenvalue of $J$ and let $\lambda_{max}$ be its real part. The return time $T_R = 1 / |\lambda_{max}|$. For a steady state to be stable, $\lambda_{max}$ is $< 0$. The steady state progressively loses stability as the bifurcation point is approached with $\lambda_{max} \to 0$. The divergence of the return time at the bifurcation point indicates critical slowing down in the vicinity of the critical point. The dynamics of the two-gene motif are characterised by two modes: slow and fast [12,25,26]. The eigenvector $V_i$ (I = 1, 2) sets the direction in the two-dimensional state space along which

$$\delta x(t) = e^{\lambda_i t} V_i \qquad (9)$$

Substitution of equation (9) into equation (6) results in the eigenvalue equation $J_i V_i = \lambda_i V_i$, $i = 1, 2$. The eigenvectors $V_1$ and $V_2$ set the directions of slow and fast dynamics respectively in the state space since $\lambda_1$ is $> \lambda_2$ (less negative) so that the exp $(\lambda_2 t)$ term in equation (8) decays faster. In the limit of large times, the trajectories in the state space approach the stable steady state tangential to the slow eigendirection. Figure 4(i) shows some of these trajectories computed by solving equations (1) and (2). The parameter values used are $a = 5.0$, $b = 2.0$, $k = 1.0$, $g = 0.5$ and $S = 2.0$. Figure 4(ii) shows that the direction of slow relaxation is also the direction of large fluctuations (variance). The distribution of the $(x_1, x_2)$ values has been obtained by solving the Langevin equations [12]. The variance increases as the bifurcation point is approached. The covariance matrix $CVM$ in the case of the two-gene motif has the form

$$CVM = \begin{bmatrix} \langle(\delta x_1)^2\rangle & \langle\delta x_1 \delta x_2\rangle \\ \langle\delta x_1 \delta x_2\rangle & \langle(\delta x_2)^2\rangle \end{bmatrix} \qquad (10)$$

with the diagonal elements representing the variances. Near criticality, the CVM is given by

$$CVM = \sigma^2 \begin{bmatrix} 1 & \dfrac{\beta_1}{r_{12}} \\ \dfrac{\beta_1}{r_{12}} & \left(\dfrac{\beta_1}{r_{12}}\right)^2 \end{bmatrix} \qquad (11)$$

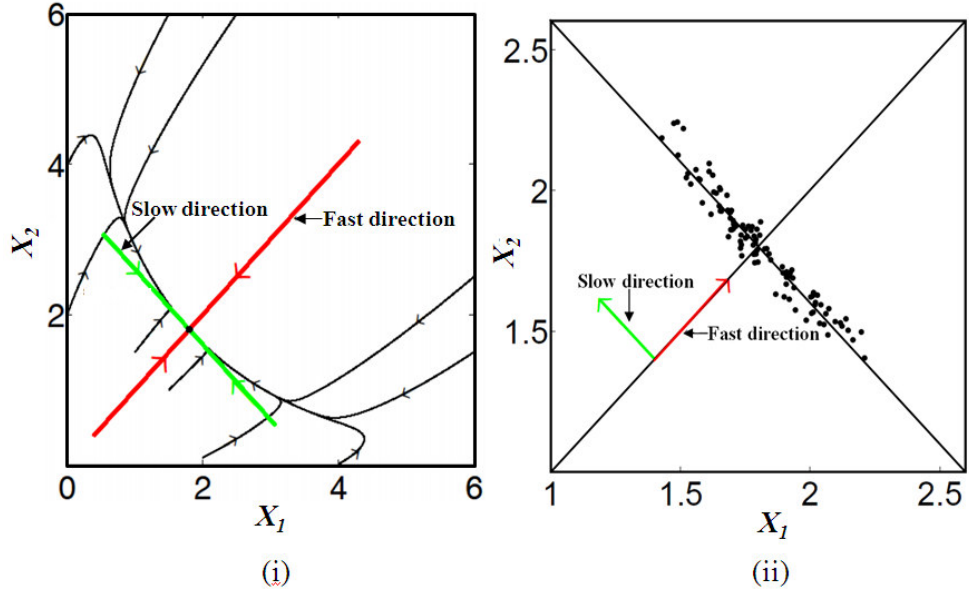

Figure 4. (i) Trajectories in state space become tangential to the slow eigendirection in the limit of large time. (ii) Variance is large (small) along the slow (fast) eigendirection.

The eigenvector of the *CVM* with the larger eigenvalue, termed the first principal component in the standard principal component analysis (PCA) [27], sets the direction in state space along which the data variation is maximal. Close to criticality, i.e., when $\lambda_1 \to 0$, the direction becomes the same as that corresponding to the slow mode of the dynamics. In the case of several genes, the state space is high-dimensional and the PCA provides an efficient tool for analyzing the data. The PCA determines new variables, termed the principal components (PCs), which are linear combinations of the original state space variables. The PCs can be shown to be the normalized eigenvectors of the covariance matrix with the eigenvectors being ordered by the magnitudes of the eigenvalues. The eigenvector corresponding to the largest eigenvalue defines the first PC which sets the direction of maximum variation in the state space. The second PC accounts for the dominant part of the remaining variation and so on. Thus, one needs to consider only the first few PCs to retain most of the variation present in the original data set. If the original data constitute an *m*-dimensional state space and one retains the first *q* PCs ($q < m$), there is an effective dimensional reduction in the analysis of the data without significant loss in information.

The experiment by Chang et al. [23] on blood cell differentiation in mice provides evidence of both slow kinetics and broad heterogeneity (large variance). The distribution of the Sca-1 marker protein levels in the undifferentiated cell population exhibits considerable heterogeneity. In the experiment, the clonal cell population was sorted into three subpopulations with low, medium and high Sca-1 levels respectively. The distributions of the Sca-1 levels in the subpopulations were found to revert back to the distribution in the parental cell population through slow reaction kinetics. In Ref. [12], the conjecture was put forward, based on theoretical analysis, that the experimentally observed features are a manifestation of criticality, i.e., the proximity of the differentiating cell population to a bifurcation (critical) point. In a subsequent study [28], Ridden et al. analyzed the expression fluctuations of the marker protein Sca-1 in terms of a simple stochastic model (model fit to experimental data was very good) and showed that the computed distribution in the Sca-1 levels lies close to the critical lines separating bistable and monostable regimes and also the maximum entropy distribution. The latter result is consistent with Jaynes' Maximum Entropy Principle [29] according to which the equilibrium distribution maximizes entropy subject to the constraints imposed on the system. The removal of a constraint increases the entropy of a system. The entropy of a gas confined by a piston in a cylinder is increased when the constraint (piston) is removed and the gas is able to expand freely. In terms of the regulatory constraints of a genetic network, the undifferentiated pluripotent cell population has entropy higher than that of the more tightly regulated differentiated cell populations. Since entropy is a measure of variability, the pluripotent cell population is expected to be more heterogeneous. This is borne out by the steady state distributions exhibited in figure 3.

Recently, a large number of studies have been carried out on the early signatures of sudden regime shifts in systems as diverse as ecosystems, financial markets, population biology and complex diseases [30,31]. Similar studies on the signatures of regime shifts in gene expression dynamics have been carried out in Refs. [12,32]. Regime shifts mostly occur at bifurcation points or may be noise-induced. The early signatures of approaching regime shifts include the critical slowing down and its associated effects, namely, a rising variance and lag-1 autocorrelation function, and an increasingly skewed steady state probability distribution. These quantities are experimentally measurable and provide advanced knowledge of the regime shifts occurring at bifurcation (critical) points. In the case of a random variable $x$, the variance of $x$ is expressed as Var $(x) = <x^2> - <x>^2$ where $<...>$ represents the mean value over an ensemble of cells. The lag-1 autocorrelation function, $ACF$, is given by

$$ACF = \frac{\langle x(t+1)x(t)\rangle - \langle x(t+1)\rangle\langle x(t)\rangle}{\sqrt{Var\big(x(t+1)\big)Var(x(t))}} \qquad (12)$$

The critical slowing down close to a bifurcation point implies that the system's intrinsic rates of change are lowered. The fluctuations in a state variable thus have a greater propensity for accumulation (rising variance) as the bifurcation point is approached. Similarly, the state of the system at time $t+1$ closely resembles the state at time $t$ giving rise to increased lag-1 autocorrelation which reaches its maximum value at the bifurcation point. The quantities providing early signatures have been computed in Ref. [12] in the case of the two-gene motif and their behaviour is along expected lines. Another signature of criticality was obtained in terms of a perfect anticorrelation between the two variables $x_1$ and $x_2$. The correlation coefficient, $CCF$ (also known as Pearson's correlation coefficient), is given by the normalized covariance,

$$CCF = \frac{\langle \delta x_1 \delta x_2 \rangle}{\sqrt{\langle(\delta x_1)^2\rangle\langle(\delta x_2)^2\rangle}} \qquad (13)$$

From equations (10) and (11), the *CCF* = -1 since the coefficient $r_{12}$ = -1 (repressive influence). Experiments detecting the critical slowing down are difficult to carry out but some recent experiments demonstrate the feature in specific biological systems [33,34]. Quantities like the variance and the lag-1 autocorrelation function are measurable using single cell methodologies like flow-cytometry and time-lapse fluorescence spectroscopy [35,36]. The proposal made in [12] that the proximity of a bifurcation point confers the experimentally observed features of broad heterogeneity and slow relaxation kinetics on the cell population can be tested in the experimental measurements mentioned. The signatures would be missing in the case of pure noise-driven transitions. A very recent experimental study by Mojtahedi et al. [24] provides strong support to the idea of critical state transitions. In the experiment it was shown that the destabilization of a high-dimensional attractor state brings about the exit of cells from the pluripotent progenitor state paving the way for cell differentiation into specific lineages. A low-dimensional representation of the data in high-dimensional state space was obtained using the PCA. Single-cell monitoring of the gene expression profiles of cell populations in high-dimensional state space was further carried out and a new index $I_C$ computed which provides the signature for an approaching critical point transition. $I_C$ is defined to be

$$I_C(t) = \frac{\langle |R(g_i, g_j)| \rangle}{\langle R(S^k, S^l) \rangle} \qquad (14)$$

where *R* is the Pearson's correlation coefficient between the cell-state vectors $S^k$ and $S^l$ or between the gene vectors $g_i$ and $g_j$. The brackets <….> denote the average of *R* over all pairs of state or gene vectors associated with a cell population. Let *n* be the number of cells in the population and *m* be the number of genes the expressions of which are monitored in each individual cell. The state vector $S^k$ of the *k*-th cell at time *t* is defined by the protein concentrations, $x_i$'s, *i = 1,…,m*. The gene vector $g_i$ is defined by the expression levels (protein concentrations) of the *i*-th gene in each of the *n* cells. The magnitude of $I_C$ increases as the bifurcation point is approached and attains its maximum value at the bifurcation point itself.

The notion of criticality in biological systems is not new and there have been suggestions that biological networks like neuronal and genetic networks operate at criticality with the dual functional advantages of robustness and flexibility [4,37]. The critical state separates ordered from disordered phases, the ordered phase is resistant to perturbations, i.e., robust whereas the disordered phase favours variable (flexible) responses. A recent study has analysed the experimental gene expression data of a gap-gene network in a developing fruit fly embryo and has identified signatures that the genetic network is tuned to criticality [25]. The gap-gene network has four main gap genes. There are, however, regions in the expression profile with simpler dynamics involving a pair of mutually repressive gap genes. There is experimental evidence of strong anticorrelation of magnitude – 1 between the fluctuations of the two protein levels. The other signatures of criticality include slow dynamics along some directions in state space with large variances of the expression data in these directions, long-range spatial correlations of expression fluctuations in the embryo and a non-Gaussian distribution of these fluctuations. In both the cases of the fruit fly embryo at the developmental stage and blood cell differentiation, the two-gene motif of mutual repression of gene expression is a key component of the regulatory genetic network. The mathematical analysis of criticality is similar in the two cases and forms the basis for identifying criticality in more general gene regulatory networks [38,39]. In the cases of gene regulatory networks, dynamical regime changes mostly occur through bifurcations. There is a large body of literature on generic signatures of approaching bifurcation points in diverse systems [30,31] conferring a universal character on criticality associated with such systems. The search for such signatures in the dynamical behavior of gene networks is in its early stages with only a few theoretical and experimental studies carried out to date [12, 18, 24, 25]. In this context, we mention a recent experimental result [40] which may be interpreted in terms of an approaching critical point. The transcription factor Yan plays a key role in maintaining the Drosophila (fruit fly) eye cells in a multipotent state. Yan is a core component of the gene regulatory network which control the time and place of the multipotent cells into several differentiated lineages. A sharp spike in the Yan expression noise was observed in the experiment coinciding with the early stages of the transition of progenitor cells to differentiated states. If the transition occurs via a bifurcation, the increasing noise may be attributed to a rising variance associated with criticality. The study of Mojtahedi et al. [24] acquires significance for providing experimental support to the criticality conjecture in cell differentiation [12]. The generality of the idea needs to be tested in different cell lines. On a more fundamental level, one needs to understand the advantages arising from the dynamical tuning of living systems to be near criticality [41].

## Acknowledgements

IB acknowledges the support by CSIR, India, vide sanction Lett. No. 21(0956)/13-EMR-II dated 28.04.2014. MP acknowledges support from Bose Institute, India for carrying out the research. The Authors thank Achintya Singha for his help in preparing the manuscript.

## Glossary: Basic Terminologies of Nonlinear Dynamics

Dynamical system: a system the state of which evolves as a function of time.

State at time t: defined by the magnitudes of the key variables at time t.

State space: the space of all states. For an N-variable system the state space is N-dimensional with one coordinate axis for each variable.

Time evolution: described by differential rate equations, one equation for each variable.

Trajectory in state space: each state is defined by a single point in state space. A trajectory is obtained by joining the points representing the states of the system at successive time intervals. A knowledge of the state at time is obtained by solving the differential rate equations.

Steady state: a state in which all rates of change are zero, defines the fixed point of the dynamics.

Stability of steady state: a steady state is stable (unstable) if the system comes back to (goes away from) the steady state after being weakly perturbed from it.

Bistability: two stable steady states are possible for the same parameter values.

Basin of attraction of a stable steady state: defined by a region in state space. All trajectories in this region end up at the fixed point representing the stable steady state.

Attractors of dynamics: trajectories in state space end up in the attractors which include both fixed points and limit cycles. In the latter case, the attractor is a cycle of states repeatedly traversed by the system indicative of periodic motion.

Bifurcation: occurs at specific parameter values at which there is a change in the dynamical regime, involves changes in the number and/or the stability properties of steady state solutions of the differential rate equations.

**References**

[1]   Moris N, Pina C and Arias A M 2016 Nat. Rev. Genet. 17 693

[2]   Zhou J X and Huang S 2011 Trends Genetics 27 55

[3]   MacArthur B D, Ma'ayan A and Lemischka I R 2009 Nat. Rev. Mol. Cell Biol. 10 672

[4]   Kauffman S A 1993 The Origins of Order: Self-Organization and Selection in Evolution (New York: Oxford University Press)

[5]   Kauffman S A 1969 J. Theor. Biol. 22 437

[6]   Huang S, Eichler G, Bar-Yam Y and Ingber D E 2005 Phys. Rev. Lett. 94 128701

[7]  Ferrell J E 2012 Curr. Biol. 22 R458

[8]  Enver T, Pera M, Peterson C and Andrews P W 2009 Cell Stem Cell 4 387

[9]  Wang J, Xu L, Wang E and Huang S 2010 Biophys. J. 99 29

[10] Foster D V, Foster J G, Huang S and Kauffman S A 2009 J. Theoret. Biol. 260 589

[11] Heinäniemi M et al. 2013 Nat. Methods 10 577

[12] Pal M, Ghosh S and Bose I 2015 Phys. Biol. 12, 016001

[13] Cahan P and Daley G Q 2013 Nat. Rev. Mol. Cell Biol. 14 357

[14] Huang S 2009 Development 136 3853

[15] Kærn M, Elston T C, Blake W J and Collins J J 2005 Nat. Rev. Genet. 6 451

[16] Raj A and van Oudenaarden A 2008 Cell 135 216

[17] Raj A and van Oudenaarden A 2009 Ann. Rev. Biophys. 38 255

[18] Garcia-Ojalvo J and Arias A M 2012 Curr. Opin. Genet. Development 22 619

[19] MacArthur B D and Lemischka I R 2013 Cell 154 484

[20] Zhou J X, Aliyu M D S, Aurell E and Huang S J. 2012 Royal Soc. Interface 9 3539

[21] van Kampen N G 1992 Stochastic Processes in Physics and Chemistry (Amsterdam: Elsevier)

[22] Fox R F, Gatland I R, Roy R and Vemuri G 1988 Phys. Rev. A 38 5938

[23] Chang H H, Hemberg M, Barahona M, Ingber D E and Huang S 2008 Nature 453 544

[24] Mojtahedi M et al. 2016 bioRxiv 041541

[25] Krotov D, Dubuis J O, Gregor T and Bialek w 2014 Proc. Natl. Acad. Sci. USA 111 3638

[26] Strogatz S H 1994 Nonlinear Dynamics and Chaos: with Applications to

Physics, Biology, Chemistry and Engineering (Reading, MA: Addison-Wesley)

[27] Ringnér M 2008 Nat. Biotech. 26 303

[28] Ridden S J, Chang H H, Zygalakis K C and MacArthur B D 2015 Phys. Rev. Lett. 115 208103

[29] Jaynes E T 1957 Phys. Rev. 106 620

[30] Scheffer M et al. 2009 Nature 461 53

[31] Scheffer M et al. 2012 Science 338 344

[32] Pal M, Pal A K, Ghosh S and Bose I 2013 Eur. Phys. J. E 36 123

[33] Dai L, Vorselen D, Korolev K S and Gore J 2012 Science 336 1175

[34] Veraart A J, Faassen E J, Dakos V, van Nes E H, Lüring M and Scheffer M 2012 Nature 481 357

[35] Weinberger L S, Dar R D and Simpsn M L 2008 Nat. Genetics 40 466

[36] Kaufmann B B and van Oudenaarden A 2007 Curr. Opin. Genet. Dev. 17 107

[37] Gros C 2008 Complex and Adative Dynamica Systems; a Primer (New York: Springer)

[38] Valverde S, Ohse S, Turalska M, West B J and Garcia-Ojalvo J 2015 Front. Physiol. 6 127

[39] Podolskiy D et al. 2015, arXiv: 1502.04307 v1 [q-bio.MN]

[40] Peláez N et al. 2015 eLife 4 e08924

[41] Hidalgo J, Grilli J, Suweis S, Muñoz M A, Banavar J R and Maritan A 2014 Proc. Natl. Acad. Sci. USA 111 10095